**Full citation**: Han, S., Sinha, R., & Lowe, A. (2020) Assessing Support for Industry Standards in Reference Medical Software Architectures, in *Proceedings of the 46th Annual Conference of the IEEE Industrial Electronics Society (IECON2020)*. Singapore, IEEE Computer Society Press, pp.3403-3407. doi: 10.1109/IECON43393.2020.9255309.

# Assessing Support for Industry Standards in Reference Medical Software Architectures

Shihui Han
T & Software Engineering
Auckland University of Technology
Auckland, New Zealand
wkq7224@autuni.ac.nz

Roopak Sinha
IT & Software Engineering
Auckland University of Technology
Auckland, New Zealand
rsinha@aut.ac.nz
ORCID: 0000-0001-9486-7833

Andrew Lowe
Mechanical Engineering
Auckland University of Technology
Auckland, New Zealand
andrew.lowe@aut.ac.nz

**Abstract**

*Industrial standards for developing medical device software provide requirements that conforming devices must meet. A number of reference software architectures have been proposed to develop such software. The ISO/IEC 25010:2011 family of standards provides a comprehensive software product quality model, including characteristics that are highly desirable in medical devices. Furthermore, frameworks like 4+1 Views provide a robust framework to develop the software architecture or high level design for any software, including for medical devices. However, the alignment between industrial standards and reference architectures for medical device software, on one hand, and ISO/IEC 25010:2011 and 4+1 Views, on the other, is not well understood. This paper aims to explore how ISO/IEC 25010:2011 and 4+1 Views are supported by current standards, namely ISO 13485:2016, ISO 14971:2012, IEC 62304:2006 and IEC 62366:2015, and current reference architectures for medical device software. We classified requirements from medical devices standards into qualities from ISO/IEC 25010:2011 and architectural views from 4+1 Views. A systematic literature review (SLR) method was followed to review current references software architectures and a mapping of their support for the identified ISO/IEC 25010:2011 qualities in the previous step was carried out. Our results show that ISO/IEC 25010:2011 qualities like functional suitability, portability, maintainability, usability, security, reliability and compatibility are highly emphasised in medical device standards. Furthermore, we show that current reference architectures only partially support these qualities. This paper can help medical device developers identify focus areas for developing standards-compliant software. A wider study involving under-development medical devices can help improve the accuracy of our findings in the future.*

## 1. INTRODUCTION

There is increasing interest in the field of the design and development of medical devices. Many recent research papers have addressed standards related to medical devices, but *there is a gap in that current research has not considered improvements in the software architecture to create better support for standard-based requirements.*

*The software architecture of a complex system determines its ability to exhibit desired qualities, captured as quality attributes [1]*. ISO/IEC 25010:2011 presents a product quality model that organizes software quality attributes into 8 characteristics and 31 sub-characteristics, which help in developing software projects, software engineering specification and evaluating non-functional requirements [2]. Desired qualities and primary features are key inputs to the development of a software architecture, which involves early and high-level design and risk assessment for ensuring that the software being built will exhibit the desired qualities. The 4+1 Views model [3] is an industry-standard framework for building the architecture of a software-intensive system. It can present different aspects of the architecture via four architectural views: *logical, process, development* and *physical*. When using an object-oriented design approach, the logical view describes the object model of the design. The process view depicts the concurrency, synchronization and runtime aspects of the system. The physical view characterizes the mapping of the software onto hardware. Finally, the development view represents the organisation of the software within its development environment [4].

To ensure safety, the design and development of medical device software should comply with medical standards. IEC 62304:2006 creates a model of planning, requirements gathering, design, implementation, verification, integration testing, system testing, and publishing activities. Moreover, risk management, configuration management, and issue tracking have been extended throughout the life cycle. Risk management should follow the ISO 14971:2012 risk management standard for medical devices. Also, ISO 13485:2016 (an extended version of the ISO 9001 quality management standard for medical devices) is used for quality management [5]. The IEC 62366:2015 standard allows the integration of usability engineering processes (UEP) in their medical device design and development [6].

In this paper, we focus on the following research questions:

**RQ1** To what extent do requirements from ISO 13485:2016, ISO 14971:2012, IEC 62304:2006 and IEC 62366:2015 relate to specific ISO/IEC 25010:2011 product quality characteristics and views in the 4+1 Views framework?

**RQ2** To what extent do current reference software architectures for developing medical device software support the qualities and views identified in RQ1?

Medical standards contribute towards the creation of robust devices that exhibit desired qualities from the ISO/IEC 25010:2011 product quality model. Standards like ISO 13485:2016 provide standardized processes for creating a medical device and therefore relate better with the development view in 4+1 Views. Quality management as per ISO 13485:2016 links equally well to both the ISO/IEC 25010:2011 product quality model and several views in 4+1 Views. Mapping medical standards to both these frameworks therefore provides a holistic picture of how developers can undertake standards-driven design of medical devices.

For answering RQ1, we perform a mapping of requirements from medical standards to qualities in ISO/IEC 25010:2011 and views in 4+1 Views (Sec. 3). For RQ2, we conduct a systematic literature review, discussed in Sec. 2, to identify five reference architectures for medical software: service-oriented medical device architecture [7], fuzzy-based modular [8], intrinsically secure, open, and safe cyber-physically enabled, life-critical essential services architectures [9], sensor information systems for active retirees and assisted living [10], and model-based systems engineering [11]. We then evaluate the extent to which these architectures support each of the views in 4+1 Views, presented as a mapping in Sec. 4.

## 2. SYSTEMATIC LITERATURE REVIEW PROCESS

The primary research method used in this paper is the Systematic Literature Review (SLR) [12], which was used to identify and analyse available primary studies to answer the two research questions. Fig. 1 illustrates the SLR process followed for this research.

The search terms for this SLR included: medical device software architecture, medical IoT architecture, medical internet of things architecture, IEC 62304, ISO 14971, ISO 13485, and IEC 62366. Several combinations of these terms were used to search peer-reviewed research databases Springer, Scopus, Science Direct, and IEEEXplore, with Google Scholar used to ensure completeness. The search string was refined to the following final form: `(medical device software architecture) OR (medical IoT architecture) OR (medical`

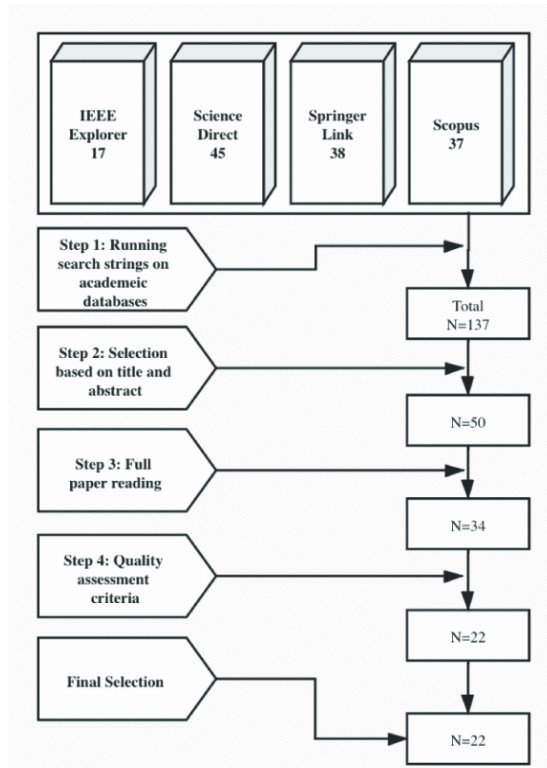

Fig. 1. the process of systematic literature review

`internet of things architecture) OR (IEC 62304 OR ISO 14971 OR ISO 13485 OR IEC 62366).`

A total of 137 studies were identified after the initial search. Inclusion and exclusion criteria were chosen to provide a basis for finding the most relevant research on software development and its related areas [13]. For this study, we included each study only after a thorough manual analysis of the title and abstract, which reduced the number of primary studies to 50. These papers were then read fully and another 16 were excluded due to their lack or relevance to our research questions. Quality assessment criteria, such as excluding non peer-reviewed research or duplicated articles provided us with 22 studies which were used.

For data extraction and synthesis, a simple tabular form was chosen to show the relationships between the requirements from standards and elements of ISO/IEC 25010:2011 and 4+1 Views. These tables are reported in subsequent sections and help answer the research questions at hand.

## 3. MAPPING OF MEDICAL DEVICE STANDARDS TO SOFTWARE ARCHITECTURE VIEWS

We created four tables (Tab. I, Tab. II, Tab. III, Tab. IV) to map requirements from each of the four medical standards we studied to the characteristics and sub characteristics in ISO/IEC 25010:2011 and the views in 4+1 Views. The column R[number] for each table shows the number of requirements that were identified for each specific ISO/IEC 25010:2011 (sub)-characteristic, and the View column shows the specific architectural view from 4+1 Views that the identified (sub)-characteristic related to.

Table I. MAPPING OF ISO 13485 REQUIREMENTS TO ISO 25010 SUB-CHARACTERISTICS AND 4+1 VIEWS

| ISO 25010 Sub-characteristic | R[num] | View |
|---|---|---|
| Security-Integrity | 1 | Process |
| Maintainability-Modifiability | 3 | Development |
| Functional suitability-Functional appropriateness | 124 | Logical |
| Portability-Installability | 3 | Physical |
| Usability-Appropriateness recognizability | 2 | Logical |

Table II. MAPPING OF ISO 14971 REQUIREMENTS TO ISO 25010 SUB-CHARACTERISTICS AND 4+1 VIEWS

| ISO 25010 Sub-characteristic | R[num] | View |
|---|---|---|
| Maintainability-Modifiability | 4 | Logical |
| Functional suitability-Functional appropriateness | 39 | Logical |
| Reliability-Fault tolerance | 1 | Process |

Table III. MAPPING OF IEC 62304 REQUIREMENTS TO ISO 25010 SUB-CHARACTERISTICS AND 4+1 VIEWS

| ISO 25010 Sub-characteristic | R[num] | View |
|---|---|---|
| Compatibility-Co-existence | 27 | Development |
| Security-Integrity | 1 | Process |
| Maintainability-Modifiability | 12 | Development |
| Functional suitability-Functional appropriateness | 67 | Logical |
| Usability-User interface aesthetics | 1 | Logical |
| Maintainability-Testability | 9 | Development |
| Security-Accountability | 1 | Process |
| Maintainability-Modularity | 1 | Development |
| Security-Authenticity | 6 | Process |
| Reliability-Availability | 1 | Process |

### A. ISO 13485:2016

There are 134 non-functional or quality requirements in ISO 13485:2016, most of which can be associated with functional appropriateness in functional suitability in ISO/IEC 25010:2011 and logical view in 4+1 Views. The remaining requirements are associated with security, maintainability, portability and usability in ISO/IEC 25010:2011. Tab. I shows the mapping of ISO 13485:2016 requirements to ISO/IEC 25010:2011 characteristics and views from 4+1 Views. The detailed mapping is presented in https://drive.google.com/file/d/1Q6DL-Y-VULgELOK-eYH-nCvFzYBmP9hp/view.

We consider a few examples from this mapping. Requirement R26 in the standard has the following description, *"The organization shall document procedures to define the controls needed for the identification, storage, security and integrity, retrieval, retention time and disposition of records."*. This security requirement and is mapped to the security-integrity sub-characteristic of ISO/IEC 25010:2011. Also, since integrity is a runtime concern, this requirement has been mapped to the process view. Requirement R28 in the standard has the following description, *"Records shall remain legible, readily identifiable and retrievable. Changes to a record shall remain Identifiable."*. This requirement is a maintainability requirement and is therefore mapped to maintainability-modifiability sub-characteristic of ISO/IEC 25010:2011. Also, since maintainability is development concern, we map this requirement to the development view.

### B. ISO 14971:2012

ISO 14971:2012 features 44 quality requirements. It is mainly associated with functional appropriateness in ISO/IEC 25010:2011 and the logical view in 4+1 Views. It also relates to maintainability and reliability, and the process view. Tab. II provides the mapping of requirements from ISO 14971:2012. The details of the mapping are available in:
https://drive.google.com/file/d/1Ee7sKO5fYtWLZzTR-9cZ9VqCEpyqSc-I/view.

We illustrate a few examples from this mapping. Requirement R49 in the standard has the following description, *"The manufacturer shall establish, document and maintain a system to collect and review information about the medical device or similar devices in the production and the post-production phases."*. This requirement is a maintainability requirement and is therefore mapped to maintainability-modifiability sub-characteristic of ISO/IEC 25010:2011. Also, since maintainability is a development concern, this requirement has been mapped to the development view. Requirement R6 in the standard has the following description, *"Risk management activities shall be planned. Therefore, for the particular medical device being considered, the manufacturer shall establish and document a risk management plan in accordance with the risk management process. The risk management plan shall be part of the risk management file."*. This requirement is a functional appropriateness requirement and is therefore mapped to functional suitability-functional appropriateness sub-characteristic of ISO/IEC 25010:2011. This requirement is been mapped to the logical view as it relates to functionality.

### C. IEC 62304:2006

There are 127 quality requirements in IEC 62304:2006 that relate primarily to functional appropriateness, co-existence and modifiability. These requirements relate mainly to the development and logical views in 4+1 Views. Tab. III provides the mapping of requirements from IEC 62304:2006. The details of the mapping are available in https://drive.google.com/file/d/11xFXk0Uq0Tfm62CkwGLDX4rQGa s8LUt/view.

We consider a few examples from this mapping. Requirement R28 in the standard has the following description, *"The MANUFACTURER shall include or reference in the software development plan, a plan to integrate the SOFTWARE ITEMS (including soup) and perform testing during integration.[Class B.C]"*. This requirement is a compatibility requirement and is therefore mapped to compatibility-co-existence sub-characteristic of ISO/IEC 25010:2011.This requirement is mapped to the development view. Requirement R110 in the standard has the following description, *"The MANUFACTURER shall EVALUATE and approve CHANGE REQUESTS which modify released MEDICAL DEVICE SOFTWARE. [Class A, B, C]"*. This requirement is a maintainability/modifiability requirement and is therefore mapped to maintainability-modifiability sub-characteristic of ISO/IEC 25010:2011 and to the development view.

### D. IEC 62366:2015

The last mapping table contains 54 quality requirements from IEC 62366:2015, more than two-thirds of which can be linked to the user interface aesthetics in usability in ISO/IEC 25010:2011 and the logical view. The remaining requirements are linked to the maintainability, reliability and functional suitability in ISO/IEC 25010:2011 and the process view and development view in 4+1 Views. Tab. IV provides the mapping of requirements from IEC 62366:2015. The details of the mapping are available in https://drive.google.com/file/d/1b6CpCjjRcIKb9Pp8rlGU1uSyjs3rJ5fF/view.

Table IV. MAPPING OF IEC 62366 REQUIREMENTS TO ISO 25010 SUB-CHARACTERISTICS AND 4+1 VIEWS

| ISO 25010 Sub-characteristic | R[num] | View |
|---|---|---|
| Usability-User interface aesthetics | 41 | Logical |
| Functional suitability-Functional appropriateness | 7 | Logical |
| Maintainability-Testability | 2 | Development |
| Reliability-Maturity | 4 | Development |

We consider a few examples from this mapping. Requirement R2 in the standard has the following description, *"USABILITY ENGINEERING activities for a MEDICAL DEVICE shall be planned, carried out, and documented by personnel competent on the basis of appropriate education, training, skills or Experience."*. This requirement is a usability/user interface aesthetics requirement and is therefore mapped to usability-user interface aesthetics sub-characteristic of ISO/IEC 25010:2011. Also, since usability is a concern in the logical view, this requirement has been mapped to the logical view. Requirement R33 in the standard has the following description, *"specify whether MEDICAL DEVICE-specific training is provided prior to the test and the minimum elapsed time between the training and the beginning of the test."*. This requirement is a maintainability/testability requirement and is therefore mapped to maintainability-testability sub-characteristic of ISO/IEC 25010:2011. Also, since maintainability is a concern in the development view, this requirement has been mapped to the development view.

## 4. SUPPORT FROM CURRENT ARCHITECTURES

Our SLR identified five reference architectures for constructing medical device software. These are service-oriented medical device architecture (SOMDA) [7], Fuzzy-Based modular [8], intrinsically secure, open, and safe cyber-physically enabled, life-critical essential services (ISOSCELES) architectures [9], sensor information systems for active retires and assisted living (SISARL) [10] and model-based systems engineering (MBSE) [11].

Each architecture was studied to identify its support for the various architectural views identified during the mapping reported in Sec. 3. This was done using the following process:

1) The sub-characteristics from Tab. I–Tab. IV were grouped by view-type as per 4+1 Views. As an example, 5 sub-characteristics, namely maintainability-modifiability, maintainability-testability, maintainability-modularity, compatibility-coexistence, and reliability-maturity were categorised as development view related qualities. Similarly, the process view covers 9 sub-characteristics.

2) For each view of each reference architecture, support was computed as the percentage of sub-characteristics directly targeted in the architecture over the total number of sub-characteristics identified for that view in step 1. For instance, SOMDA's process view directly supports security-confidentiality and security-integrity, which are 2 of the 9 sub-characteristics constituting the process view.

Tab. V shows the results of our mapping. None of the architectures support the physical view, mainly because devices are expected to be stand-alone. The support for various views varies between architectures but is consistently less than 50%. This highlights the need for these architectures to be adapted further to be more aligned with standards.

## 5. CONCLUSIONS

We presented a detailed mapping of four medical standards to highlight how they relate to the ISO/IEC 25010:2011 product quality model and 4+1 Views. ISO 13485:2016 mainly supports functional suitability in ISO/IEC 25010:2011 and the logical view in 4+1 Views; it also supports other characteristics including security, maintainability, portability and usability of ISO/IEC 25010:2011 and process view, development view and physical view of 4+1 Views (Tab. I). ISO 14971:2012 primarily supports functional suitability characteristics in ISO/IEC 25010:2011 and the logical view in 4+1 Views, with additional relevance to maintainability and reliability characteristics of ISO/IEC 25010:2011 and process view of 4+1 Views (Tab. II). IEC 62304:2006 relates to functional suitability, compatibility and maintainability in ISO/IEC 25010:2011 and the logical view and development view in 4+1 Views (Tab. III). IEC 14971 is focused on usability and functional appropriateness which relate to the logical view in 4+1 Views (Tab. IV). A further mapping, shown in Tab. V, highlights that current reference architectures for medical software provide only partial support for the ISO/IEC 25010:2011 characteristics identified in the previous mappings. This indicates that these architectures can be further improved to provide more direct support for the target ISO/IEC 25010:2011 characteristics and subsequently ease the burden of certifying software for compliance to medical standards.

The limitations of this study include possible human error and subjectivity in correlating requirements from standards to relevant characteristics and views in ISO/IEC 25010:2011 and 4+1 Views respectively. An independent evaluation of this mapping by a subject expert can be used to address this concern in the future. Also, only seven studies were identified to answer RQ2. While the SLR process followed was robust, the low number of studies indicates that the findings may not generalize to additional reference architectures.

Future directions for this research include identifying potential improvements to reference architectures through the inclusion of additional views or architectural tactics to better support medical standards. A similar mapping to the ISO 25010 process quality model can also be undertaken.

TABLE V. SUPPORT FOR ISO 25010 SUB-CHARACTERISTICS BY VIEW FOR EACH REFERENCE ARCHITECTURE

| Architecture | Logical | Physical | Process | Development |
|---|---|---|---|---|
| SOMDA | 0% | 0% | 22%<br><br>Unsupported:<br><br>Reliability- Maturity, Availability, Fault Tolerance, Recoverability<br><br>Security- Non-repudiation, Authenticity, Accountability | 20%<br><br>Unsupported:<br><br>Performance Efficiency- Time Behaviour, Resource Utilization, Capacity<br><br>Maintainability- Modularity, Reusability, Analysability, Modifiability, Testability |
| Fuzzy-Based Modular | 0% | 0% | 22%<br><br>Unsupported:<br><br>Reliability- Maturity, Fault Tolerance, Recoverability<br><br>Security- Confidentiality, Non-repudiation, Authenticity, Accountability | 10%<br><br>Unsupported:<br><br>Performance Efficiency- Time Behaviour, Resource Utilization, Capacity<br><br>Compatibility- Co-existence, Interoperability<br><br>Maintainability- Reusability, Analysability, Modifiability, Testability |
| ISOSCELES | 10%<br><br>Unsupported:<br>Functional Suitability- Functional Completeness, Functional Correctness, Functional Appropriateness<br>Usability-Appropriateness Recognizability, Learnability, Operability, User Error Protection, Accessibility | 0% | 22%<br><br>Unsupported:<br><br>Reliability-Maturity, Availability, Fault Tolerance, Recoverability<br><br>Security-Confidentiality, Integrity, Non-repudiation | 30%<br><br>Unsupported:<br><br>Performance Efficiency- Time Behaviour, Resource Utilization, Capacity<br><br>Maintainability-Reusability, Analysability, Modifiability, Testability |
| SISARL | 20%<br><br>Unsupported:<br>Functional Suitability- Functional Completeness, Functional Correctness, Functional Appropriateness<br><br>Usability-Appropriateness Recognizability, Learnability, Operability, User Error Protection | 0% | 0% | 40%<br><br>Unsupported:<br><br>Performance Efficiency- Time Behaviour, Resource Utilization, Capacity<br><br>Maintainability-Reusability, Analysability, Testability |
| MBSE | 10%<br><br>Unsupported:<br><br>Functional Suitability- Functional Completeness, Functional Correctness, Functional Appropriateness<br><br>Usability-Appropriateness Recognizability, Learnability, Operability, User Interface Aesthetics, Accessibility | 0% | 11%<br><br>Unsupported:<br><br>Reliability-Availability, Fault Tolerance, Recoverability<br><br>Security-Confidentiality, Integrity, Non-repudiation, Authenticity, Accountability | 40%<br><br>Unsupported:<br><br>Performance Efficiency- Time Behaviour, Resource Utilization, Capacity<br><br>Maintainability-Reusability, Analysability, Testability |

## REFERENCES

[1] L. Lundberg, J. Bosch, D. Häggander, and P.-O. Bengtsson, "Quality attributes in software architecture design," in Proceedings of the IASTED 3rd International Conference on Software Engineering and Applications. IASTED/Acta Press, 1999, pp. 353–362.

[2] I. M. S. Raharja and D. O. Siahaan, "Classification of non-functional requirements using fuzzy similarity knn based on iso/iec 25010," in 2019 12th International Conference on Information & Communication Technology and System (ICTS). IEEE, 2019, pp. 264–269.

[3] S. Meng, Z. Pan, W. Li, S. Xie, C. Liu, K. He, and H. Yang, "The "4+1 "view model on safe home system architecture," in 2010 IEEE International Conference on Software Engineering and Service Sciences. IEEE, 2010, pp. 352–355.

[4] P. B. Kruchten, "The 4+ 1 view model of architecture," IEEE software, vol. 12, no. 6, pp. 42–50, 1995.

[5] T. Laukkarinen, K. Kuusinen, and T. Mikkonen, "Devops in regulated software development: case medical devices," in 2017 IEEE/ACM 39th International Conference on Software Engineering: New Ideas and Emerging Technologies Results Track (ICSE-NIER). IEEE, 2017, pp. 15–18.

[6] S. Bras Da Costa, M.-C. Beuscart-Zéphir, J.-M. Christian Bastien, and S. Pelayo, "Usability and safety of software medical devices: need for multidisciplinary expertise to apply the iec 62366: 2007," 2015.

[7] M. Kasparick, M. Schmitz, B. Andersen, M. Rockstroh, S. Franke, S. Schlichting, F. Golatowski, and D. Timmermann, "Or. net: a service-oriented architecture for safe and dynamic medical device interoperability," Biomedical Engineering/Biomedizinische Technik, vol. 63, no. 1, pp. 11–30, 2018.

[8] C. C. Aguwa, L. Monplaisir, P. A. Sylajakumari, and R. K. Muni, "Integrated fuzzy-based modular architecture for medical device design and development," Journal of Medical Devices, vol. 4, no. 3, 2010.

[9] S. Harp, T. Carpenter, and J. Hatcliff, "A reference architecture for secure medical devices," Biomedical instrumentation & technology, vol. 52, no. 5, pp. 357–365, 2018.

[10] J. W. Liu, B. Wang, H. Liao, C. Huang, C.-S. Shih, T. Kuo, and A. Pang, "Reference architecture of intelligent appliances for the elderly," in 18th International Conference on Systems Engineering (ICSEng'05). IEEE, 2005, pp. 447–455.

[11] S. Corns, A. Thukral, and V. Thukral, "5.1. 1 parametric analysis through a model-based reference architecture for medical device development," in INCOSE International Symposium, vol. 24, no. 1. Wiley Online Library, 2014, pp. 406–417.

[12] B. Kitchenham, "Procedures for performing systematic reviews," Keele, UK, Keele University, vol. 33, no. 2004, pp. 1–26, 2004.

[13] R. Malhotra and A. Chug, "Software maintainability: Systematic literature review and current trends," International Journal of Software Engineering and Knowledge Engineering, vol. 26, no. 08, pp. 1221– 1253, 2016.